\documentclass[11pt, onecolumn, draftclsnofoot]{IEEEtran}
\usepackage{amsmath,amssymb}
\usepackage[dvips]{graphicx}
\usepackage{amsfonts}
\usepackage[mathscr]{eucal}
\usepackage{latexsym}
\usepackage{amsthm}
\usepackage{exscale}
\usepackage[mathscr]{eucal}
\usepackage{bm}
\usepackage[dvipsnames]{color}
\usepackage{cases}
\usepackage{epsfig}
\usepackage[center,small]{caption}
\usepackage{algorithm}
\usepackage{algorithmic}
\usepackage[verbose,nospace,sort]{cite}
\usepackage{tabularx}
\usepackage{multirow}
\usepackage{balance}
\usepackage{url}
\scrollmode

\begin{document}

\title{Time-Weighted Coverage of Integrated Aerial and Ground Networks for Post-Disaster Communications}
\author{ Xiaoli Xu and Yong Zeng\\
\thanks{ X. Xu, and Y. Zeng are with the National Mobile Communications Research Laboratory, Southeast University, Nanjing 210096, China. Y. Zeng is also with Purple Mountain Laboratories, Nanjing 211111, China (e-mail: \{yong\_zeng, xiaolixu\}@seu.edu.cn).}
}
\maketitle

\begin{abstract}
In this paper, we propose a new three dimensional (3D) networking architecture with integrated aerial and ground base stations (BSs) for swift post-disaster communication recovery. By exploiting their respective advantages  in terms of response time, coverage area, and operational duration, the proposed network is highly heterogeneous, consisting of sustained ground BSs, ground-vehicle mounted BSs, dropping-off BSs and flying BSs.  To reflect the importance of swift communication recovery and the dynamics of coverage area in post-disaster scenarios, we propose a new performance metric called ``\emph{time-weighted coverage}", which is an integration of the achieved communication coverage area multiplied with a weighting function over time. By choosing different weighting functions, the network deployment can be designed to achieve tradeoffs between the ``\emph{swift communication recovery}" and ``\emph{stable communication coverage}". Simulation results show that the proposed integrated aerial and ground network has high implementation flexibility and it can significantly enhance the communication coverage compared with the conventional approaches.
\end{abstract}
\vspace{-0.15in}

\section{Introduction}
In the past few decades, the frequency of natural disasters and man-made calamities have increased manifold, which cause heavy loss of lives and destruction of properties\cite{Disaster}. Post-disaster communication is of paramount importance for life saving. However, the ground communication infrastructure such as cellular base stations (BSs) may be destroyed or becoming malfunctioning during catastrophic disasters. Conventional emergency communication recovery usually relies on satellite, emergency communication vehicles and Walkie-Talkie \cite{Kishorbhai2017,Baldini2014,Miranda2016,Pervez2018}. Unfortunately, popular communication terminals, such as the cell phones and computers, may not communicate with satellite directly, unless extendable antennas are added. Furthermore, many catastrophic disasters such as high-magnitude earthquake and chemical explosions also usually cause severe road damages, which hinders emergency  ground communication vehicles entering into the disaster area. Walkie-Talkie also has the limitation of short communication range and its popularity is far less than cell phones.

Recently, unmanned aerial vehicles (UAV)-aided communications have received significant research attentions, and one of the important application scenario is disaster relief \cite{Zeng2016,Azari2018,Zhao2019,Panda2019,Zeng2019}.  Compared with the conventional emergency communications, UAV-aided emergency communications can be more swiftly and flexibly deployed, without being constrained by ground traffic conditions. Moreover, they usually have better channels  with ground terminals due to the high likelihood of line-of-sight (LoS) links. However, UAV-aided communications also have a critical limitation, namely the short operation time, due to the limited onboard energy. For example, rotary-wing UAVs in the market typically have the maximum  endurance of about 30 minutes. Though fixed-wing UAVs usually have longer endurance, they cannot hover at a fixed point and are more demanding  for takeoff and landing.

\begin{figure}[htb]
\centering
\includegraphics[scale=0.5]{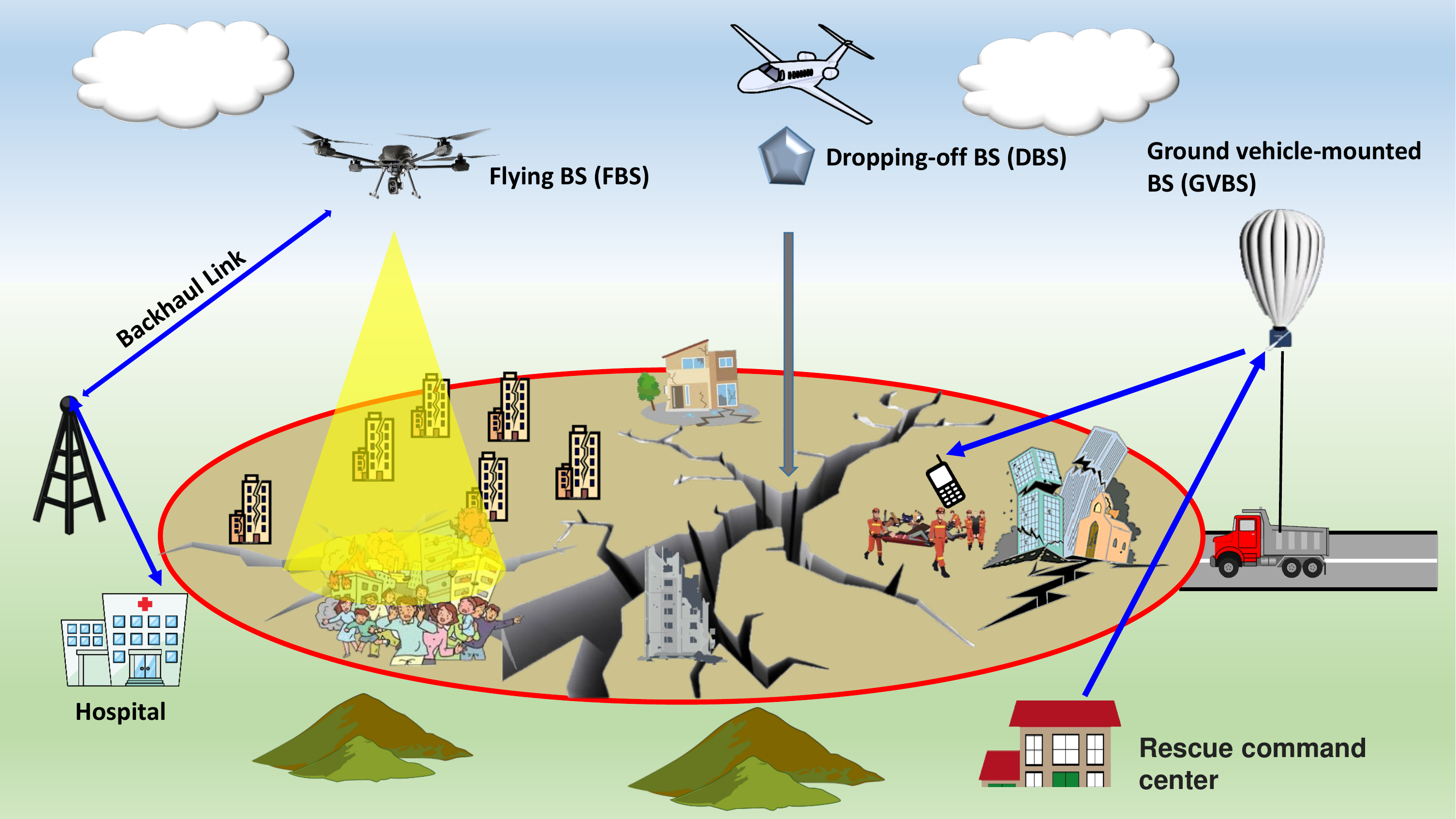}
\caption{3D networking for post-disaster communication recovery.}
\label{F:System}
\end{figure}

To address the above issues, in this paper, we propose a new three dimensional (3D) networking architecture with  integrated aerial and ground BSs for swift post-disaster communication recovery, which exploits the advantages of various types of ground and aerial BSs in terms of response time, coverage area, operation duration, etc.  As illustrated in Fig.~\ref{F:System}, the proposed 3D emergency communication network includes the following types of BSs:
\begin{itemize}
\item{\emph{Terrestrial BS (TBS)}: the remaining ground cellular BSs that are sustained during the disaster.}
\item{\emph{Ground vehicle-mounted BS (GVBS)}: the BSs carried by emergency ground communication vehicles.}
\item{\emph{Flying BS (FBS)}: compact BSs mounted by manned or unmanned aircrafts.}
\item{\emph{Dropping-off BS (DBS)}: BSs carried by manned or unmanned aircrafts to desired locations and then dropped off to the ground for communication recovery.}
\end{itemize}

Note that the different types of BSs in the proposed network are complementary with each other as they have different characteristics. For instance, the sustained TBSs typically have stable power supply and established backhaul connections, which can provide communication service immediately after the disaster. However, the number of sustained TBSs and their locations are non-predictable, depending on the type and severity of the disaster. On the other hand, GVBS, FBS and DBS have different response time, determined by the travelling distance and vehicle/aircrafts' speed. FBS and DBS usually can be deployed with much higher flexibility than GVBS, but they typically have limited endurance. Meanwhile, GVBS may have stable power supply from the vehicle and may establish backhaul connection with the satellite via extended antennas carried by vehicles. Both DBS and FBS have high deployment flexibility. However, FBS usually has larger coverage radius than DBS due to the higher altitude, while DBS usually has longer endurance.

\begin{figure}[htb]
\centering
\includegraphics[scale=0.7]{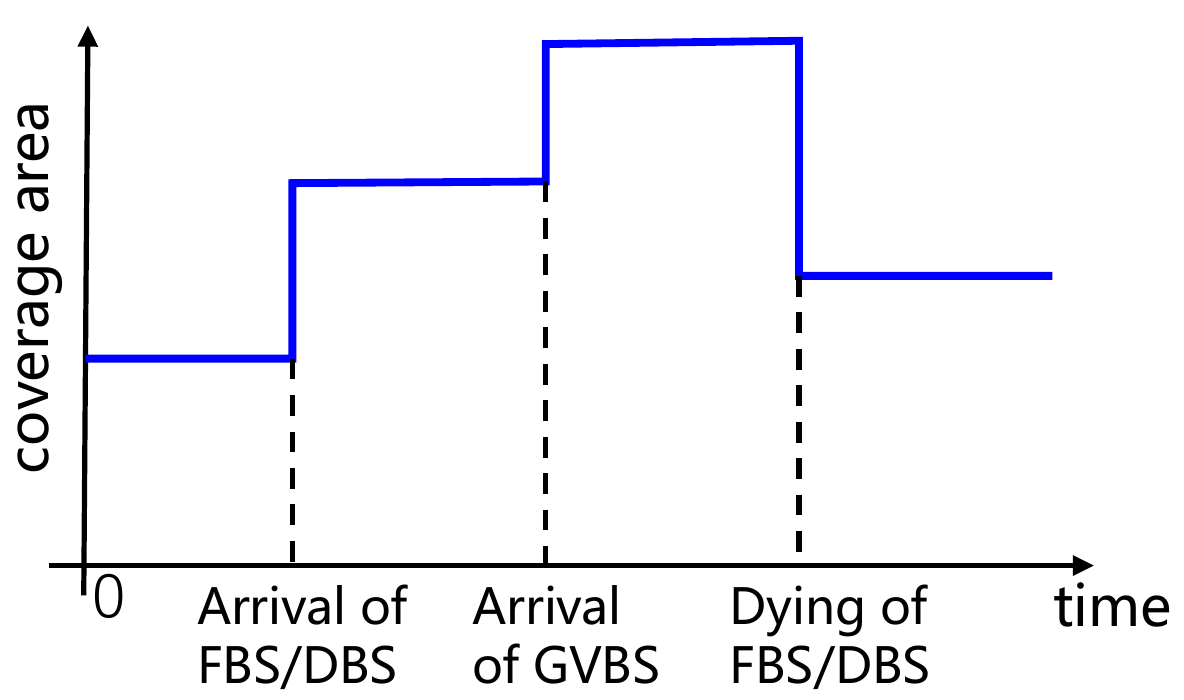}
\caption{Coverage dynamics of post-disaster communication recovery.}
\label{F:coverage}
\end{figure}

In this paper,  we investigate the joint deployment of GVBS, FBS and DBS with given locations of sustained TBS. Different from the conventional networking plan in cellular networks, the deployment design for the considered problem includes the dispatch time of vehicles/aircrafts and the designated locations of GVBS/FBS/DBS. Furthermore, the achievable coverage area during the deployment phase is highly dynamic due to the arrival of new BSs and/or departure of existing ones, as illustrated in Fig.~\ref{F:coverage}. For example, as new BSs arrive, the coverage area expands, whereas when FBS and DBS reach their limited endurance, the coverage area shrinks.
\vspace{-0.1in}

To reflect the paramount importance of \emph{swift} communication recovery, as well as the coverage dynamics shown above, we propose a new performance metric called ``\emph{time-weighted coverage}", which is the integration of coverage area multiplied with a weighting function over time. By choosing different weighting functions, the network can be designed to achieve different tradeoffs between the ``\emph{swift communication recovery}" and ``\emph{stable communication coverage}". As it is quite challenging to obtain a closed-form expression for the coverage dynamics, we formulate the network deployment problem  as black-box optimization and solve it by a two-stage algorithm. First, the GVBS deployment is optimized without considering the coupled effect of FBS and DBS. Then, we optimize the FBS/DBS deployment with the obtained GVBS deployment using genetic algorithm (GA). Our work is fundamentally different from the existing works on UAV deployment\cite{Yaliniz2016,Zhang2018,Li2018} due to the new performance metric of time-weighted coverage that captures the highly dynamic coverage during the deployment phase. The results show that the communication coverage can be greatly enhanced by incorporating FBS/DBS, since they can be deployed flexibly to the disaster area. Furthermore, despite of limited endurance of FBS and DBS, stable coverage can be achieved  by carefully designing their dispatch time and designated locations.

\section{System Model and Problem Formulation}\label{sec:model}
We consider a circular disaster area $\mathcal A$ with radius $R$, centered at the origin. We assume that there are $M$ sustained TBSs within/near the disaster area, located at $\mathcal L_T=\{(x_m^T,y_m^T), m=1,...,M$\}. Each TBS can cover a circular region with radius $r_T$. We assume that there are $G$ GVBS near the disaster region that are available  for post-disaster communication recovery. Due to ground road constraint, each GVBS can only reach $N$ possible locations near the perimeter of disaster region, denoted as $\mathcal L_V=\{(x_n^V,y_n^V),n=1,...,N\}$. The traveling time for the $g$th GVBS to reach the $n$th location is denoted as $t_{g,n}$. The coverage radius of each GVBS is denoted by $r_V$. Furthermore, we assume that there are $U$ FBSs available for deployment that are initially located at  $(\tilde{x}_u^F,\tilde{y}_u^F), u=1,...,U$, and $K$ DBSs with initial locations $(\tilde{x}_k^D,\tilde{y}_k^D),k=1,...,K$. The maximum flying speeds of FBS and DBS are $V_F$ and $V_D$, respectively. The endurance of FBS is denoted by $T_F$.  Since DBS is usually powered by battery, it also has limited operating time, denoted by $T_D$. We assume that TBS and GVBS can connect to the core network using established backhaul links or via satellite. However, FBS and DBS must build backhaul connections via their neighboring BSs.
\begin{figure}[htb]
\centering
\includegraphics[scale=0.6]{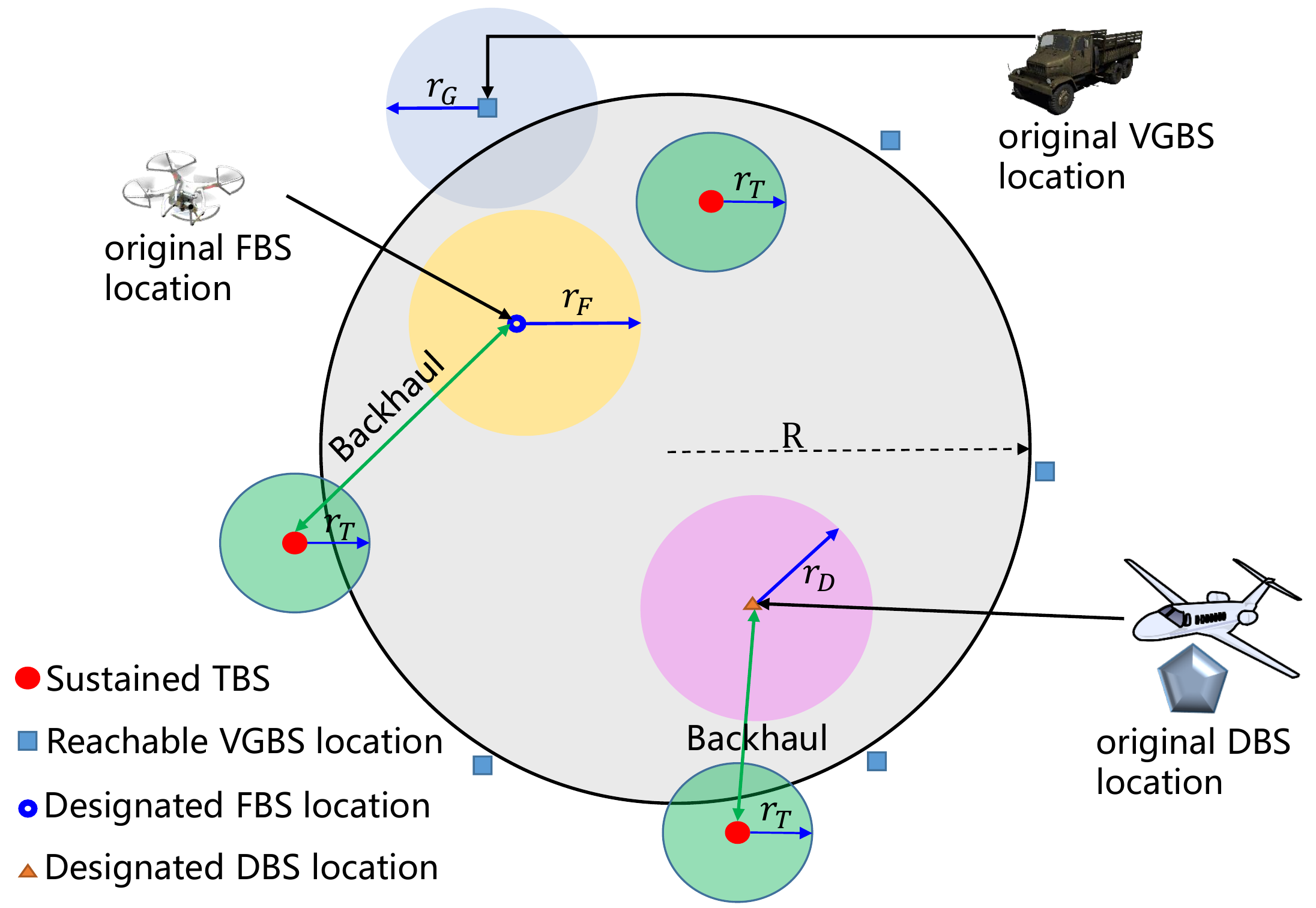}
\caption{An illustration for the deployment of the integrated aerial and terrestrial communication networks.}
\label{F:Model}
\end{figure}
\vspace{-0.2in}
\subsection{Design Variables}
There are two sets of design variables in this network deployment problem, i.e., the locations where GVBS/FBS/DBS should be placed and the time to start dispatching the vehicles/aircrafts carrying these BSs. Note that due to the limited endurance of FBS and DBS and the requirement for establishing backhaul connection before providing coverage service, FBS and DBS may not necessarily be dispatched immediately after disaster occurs. Hence, the dispatch time is also a design variable.

Since GVBS can only be deployed at a finite number of reachable locations, the placement can be represented by a set of binary variables $\{p_{g,n},g=1,...,G; n=1,...,N\}$, with
\begin{align*}
p_{g,n}=
\begin{cases}
1, & \textnormal{if the $g$th GVBS is deployed at $(x_n^V,y_n^V)$ }\\
0, & \textnormal{otherwise}
\end{cases}
\end{align*}
As each GVBS can only be deployed at one location, we have the constraint $\sum_{n}p_{g,n}\leq 1, \forall g$.  The dispatch time of GVBSs is set as $t=0$ since they are not subject to the endurance limitation.

\begin{figure}[htb]
\centering
\includegraphics[scale=0.6]{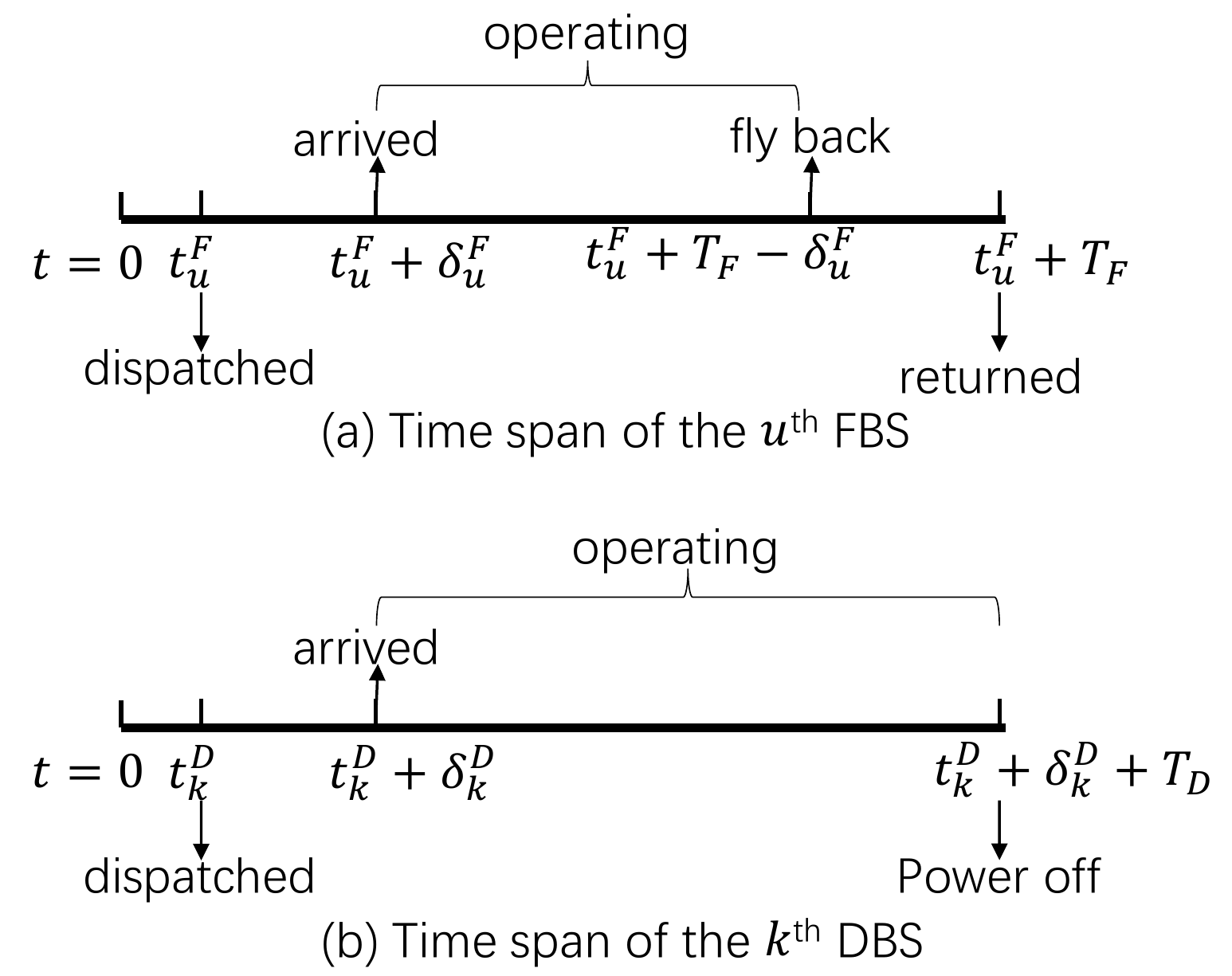}
\caption{The time span of FBS and DBS.}
\label{F:TimeSpan}
\end{figure}

For the $u$th FBS, denote its designated location by $(x_{u}^F,y_u^F)$. The time required for it to reach the designated location is
\begin{align}
\delta_{u}^F=\frac{\sqrt{(\tilde{x}_u^F-x_u^F)^2+(\tilde{y}_u^F-y_u^F)^2}}{V_F}.
\end{align}
Since the aircrafts carrying FBSs need to reserve sufficient energy for flying back to the recharge spot, the total service time is $T_F-2\delta_{u}^F$, where $T_F$ is the endurance of FBS. Due to the limited endurance, we need to choose the dispatch time of FBS carefully. Denote by $t_u^F$ the time for dispatching the $u$th FBS. The time span of the $u$th FBS is illustrated in Fig.~\ref{F:TimeSpan}(a).

Similarly, denote the designated location for the $k$th DBS by $(x_{k}^D,y_k^D)$. The time required for the aircraft to deliver the $k$th DBS is
\begin{align}
\delta_{k}^D=\frac{\sqrt{(\tilde{x}_k^D-x_k^D)^2+(\tilde{y}_k^D-y_k^D)^2}}{V_D}.
\end{align}
The aircraft carrying the $k$th DBS is dispatched at $t_k^D$. The DBS operates when it reaches the designated location and it will stop working when it runs out of the battery. The time span of the $k$th DBS is illustrated in Fig.~\ref{F:TimeSpan}(b).

In summary, the set of design variables for the integrated aerial and ground network deployment, denoted by $\mathbf\Psi$, include
\begin{itemize}
\item{The placement of GVBSs, represented by the binary variables $\{p_{g,n},g=1,...,G,n=1,...,N\}$.}
\item{The designated locations of FBSs $\mathcal L_F=\{(x_u^F,y_u^F),u=1,...,U\}$ and the corresponding dispatch time $\{t_u^F,u=1,...,U\}$.}
\item{The designated locations of DBSs $\mathcal L_D=\{(x_k^D,y_k^D),k=1,...,K\}$ and the corresponding dispatch time $\{t_k^D,k=1,...,K\}$.}
\end{itemize}

\subsection{Time-Weighted Coverage}
In this paper, we are interested in the \emph{area coverage}, i.e., the percentage of the disaster area that can be covered by the deployed BSs. The proposed network design can be easily extended to other coverage definitions, e.g., the population coverage. Note that different from the conventional network deployment for coverage maximization, the area coverage for post-disaster emergency communication is a dynamic function with time due to the arrival of new BSs and/or dying of existing ones during the deployment phase, as shown in Fig.~\ref{F:coverage}. In other words, the coverage area is a function of the deployment strategy $\mathbf\Psi$ and time $t$, and hence is denoted by $f(\mathbf\Psi,t)$.

If all the GVBSs, FBSs and DBSs are dispatched to locations close to their original locations immediately after disaster occurs, the coverage $f(\mathbf\Psi,t)$ can be boost in a short time, which leads to swift communication recovery. However, when the FBSs and DBSs reach their endurance limitations, the coverage area will shrink quickly. To achieve different trade-offs between the ``swift communication recovery" and ``stable communication coverage", we  define the time-weighted coverage as
\begin{align}
C_w=\int_{0}^{T_{th}}w(t)f(\mathbf\Psi,t)dt,\label{eq:TWC}
\end{align}
where $T_{th}$ is the time duration of interest, e.g., the ``Golden 72 hours" for life savings; $w(t)$ is the weighting function which is usually decreasing for emphasizing the importance of coverage recovery as soon as possible. In the following, we will investigate the network deployment for maximizing the time-weighted coverage in \eqref{eq:TWC}.

\subsection{Problem Formulation}\label{sec:formulation}
Denote by $\mathcal C(x,y,r)$ the circular area centered at point $(x,y)$ with radius  $r$. The TBS located at $(x_m^T,y_m^T)$ can cover the area $\mathcal C(x_m^T,y_m^T,r_T)$.  Then, the total area under the coverage of the sustained TBSs can be expressed as $\mathcal A_T\triangleq \bigcup\{\mathcal C(x_m^T,y_m^T,r_T),\forall m\}$. Hence, the coverage at $t=0$ is
\begin{align}
f(\mathbf\Psi,0)=\frac{\mathcal A_T\bigcap\mathcal A}{\mathcal A}, \label{eq:cover0}
\end{align}
where $\mathcal A$ is the whole disaster region.

For $t>0$, the area coverage is determined by the network deployment $\mathbf\Psi$. Let $\Phi_G(t)\subseteq \mathcal L_V$ be the set of locations where a GVBS has arrived at time $t$, i.e., $\forall (x_n^V,y_n^V)\in\mathcal L_V$, $(x_n^V,y_n^V)\in \Phi_G(t)$ if and only if $\exists g$ such that $p_{g,n}=1$ and $t_{g,n}<t$. Since the backhaul connections of GVBSs have been established, GVBSs are assumed to work immediately as they arrive at the designated locations. Hence, $\Phi_G(t)$ also corresponds to the set of GVBSs that actively serve users at time $t$, which cover the area
\begin{align}
\mathcal A_V(t)=\bigcup\{\mathcal C(x_n^V,y_n^V,r_V),\forall (x_n^V,y_n^V)\in\Phi_G(t)\}.
\end{align}

Further denote by $\Phi_F(t)\subseteq \mathcal L_F$ and $\Phi_D(t)\subseteq\mathcal L_D$ the set of locations with FBS and DBS arrived and actively serving users at time $t$, respectively. Note that a FBS at location $(x_u^F,y_u^F)$ is active if and only if the following conditions are met
\begin{itemize}
\item{The $u$th FBS has arrived at its designated location, i.e., $t_u^F+\delta_u^F\leq t$.}
\item{The $u$th FBS can establish backhaul connections with any other active BSs, including TBS in $\mathcal L_T$, GVBS in $\Phi_G(t)$, FBS in $\Phi_F(t)$, or DBS in $\Phi_D(t)$.  }
\end{itemize}
 Note that an FBS may become inactive, i.e., excluded from $\Phi_F(t)$, either because it leaves the location or loses the backhaul connection, e.g., due to the departure/power drain of other FBS/DBS. The area under the coverage of FBS at time $t$ is given by
 \begin{align}
 \mathcal A_F(t)=\bigcup\{\mathcal C(x_u^F,y_u^F,r_F),\forall (x_u^F,y_u^F)\in\Phi_F(t)\}.
 \end{align}

Similarly, we have $(x_k^D,y_k^D)\in\Phi_D(t)$ if and only if the $k$th DBS has arrived at this location and established the backhaul connection. The area under the coverage of DBS at time $t$ is given by
\begin{align}
\mathcal A_D(t)=\bigcup\{\mathcal C(x_k^D,y_k^D,r_D),\forall (x_k^D,y_k^D)\in\Phi_D(t)\}.
\end{align}

Hence, the total coverage at $t$ can be expressed as
\begin{align}
f(\mathbf\Psi,t)=\frac{(\mathcal A_T\bigcup\mathcal A_V(t)\bigcup\mathcal A_F(t)\bigcup\mathcal A_D(t))\bigcap\mathcal A}{\mathcal A}. \label{eq:coverage}
\end{align}

The network deployment problem is then to optimize the set of parameters $\mathbf\Psi$ so that the time-weighted coverage $C_w$ defined in \eqref{eq:TWC} is maximized. Note that $C_w$ is dependent on $\mathbf\Psi$ via \eqref{eq:cover0}-\eqref{eq:coverage}. Mathematically, we have
\begin{align}
\textbf{P1:} &\max_{\mathbf\Psi}\ C_w\\
&\textnormal{s.t., }\eqref{eq:cover0}-\eqref{eq:coverage}.
\end{align}

\section{Proposed Solution}
For a given  deployment strategy $\mathbf\Psi$, we can calculate $C_w$ by tracing the change of the network status and calculate the coverage for each state. The detailed steps are summarized in Algorithm~\ref{alg1}. However, given the deployment resource constraint, it is challenging to find the optimal network deployment $\mathbf\Psi$ that maximizes $C_w$ due to:
\begin{itemize}
\item{For more than three circles, it is difficult to derive the general expression for their union area \cite{Librino2015An}. Hence, finding the closed-form expression for $f(\mathbf\Psi,t)$ is very challenging, if not impossible.  }
\item{Problem \textbf{P1} involves both the binary variables for GVBS placement and the continuous variables for FBS/DBS deployment, which is non-convex.}
\item{The activity status of an FBS/DBS depends on the status of GVBS and other FBS/DBS. }
\end{itemize}

%To address above difficulties, we first propose to discretize the area of interest into grid points and calculate the coverage by measuring the number of covered points versus the total number of points, as shown in Fig.~\ref{F:Initial}.
%\begin{figure}[htb]
%\centering
%\includegraphics[scale=0.62]{InitialCoverage}
%\caption{The coverage area by the remaining TBS in a disaster area.}
%\label{F:Initial}
%\end{figure}

Therefore, a tractable expression for the time-weighted coverage as a function of the design variables is difficult to obtain. On the other hand, given the deployment $\mathbf\Psi$, we can calculate the time-weighted coverage via Algorithm~\ref{alg1}.  Hence, the problem \textbf{P1} can be classified as a ``black-box" optimization problem \cite{TVMeta}, which involves a set of binary variables and another set of continues variables. To tackle the above difficulties, we propose an efficient heuristic solution by decoupling the deployment of GVBS versus that of FBS and DBS. Specifically, the GVBS deployment is firstly optimized without considering the coupled effect of FBS and DBS  and then the FBS/DBS deployment is optimized with the obtained GVBS deployment using GA.

%The details of the proposed approach will be discussed in the following subsections.
\begin{algorithm}[htb]
\caption{Weighted coverage calculation with given $\mathbf\Psi$}
\label{alg1}
\begin{algorithmic}[1]
%\STATE{\textbf{Input: }$\mathcal L_T$,$\mathbf\Psi=\{\{p_{g,n},\forall g,n\},\{(x_u^F,y_u^F),t_u^F,\forall u\},\{(x_k^D,y_k^D),t_k^D,\forall k\}\}$ }
\STATE{\textbf{Input:} $\mathcal L_T,r_T,r_G,r_F,r_D,T_{th}$,$\mathbf\Psi$,$w(t)$, and the backhaul communication distance thresholds. }
\STATE{\textbf{Initialize:} $t_0=0$ }
\STATE{Arrived GVBS $\Phi_V=\emptyset$.}
\STATE{Arrived active and inactive FBS $\Phi_F,I_F=\emptyset$.}
\STATE{Arrived active and inactive DBS $\Phi_D,I_D=\emptyset$.}
\STATE{Aflag=False, Dflag=False.}
%\STATE{The start time $t_0=0$.}
\STATE{Calculate the arriving time for all GVBSs. }
\STATE{Calculate the arriving and dying time for all FBSs/DBSs.}
\STATE{Sort the above  time instances which are less than $T_{th}$ in increasing order and store them in $\mathcal T$.}
\FOR{$t_e\in\mathcal T$}\label{start0}
\STATE{Calculate the coverage $f(t,\mathbf\Psi)$ for $t\in(t_0,t_e]$ based on $\mathcal L_T,\Phi_V,\Phi_F,\Phi_D$.}
\IF{a GVBS arrives at this time instance}
\STATE{Add this GVBS into $\Phi_V$ and set Aflag=True.}
\ELSIF{a FBS/DBS arrives at this time instance}
%\STATE{Check backhaul connection of this FBS/DBS. }
\IF{backhaul connection can be established}
\STATE{Add this FBS/DBS to $\Phi_F$/$\Phi_D$ and set Aflag=True.}
\ELSE
\STATE{Add this FBS/DBS to $I_F$/$I_D$.}
\ENDIF
\ELSIF{a FBS/DBS returns at this time instance}
\IF{this FBS/DBS is inactive}
\STATE{Remove this FBS/DBS from $I_F$/$I_D$.}
\ELSE
\STATE{Remove this FBS/DBS from $\Phi_F$/$\Phi_D$ and set Dflag=True.}
\ENDIF
\ENDIF
\WHILE{Aflag=True}
\STATE{Set Aflag=False.}
\IF{any FBS/DBS in $I_F$/$I_D$ can be activated}
\STATE{Put them into $\Phi_F$/$\Phi_D$ and set Aflag=True.}
\ENDIF
\ENDWHILE
\WHILE{Dflag=True}
\STATE{Set Dflag=False.}
\IF{any FBS/DBS in $\Phi_F$/$\Phi_D$ is deactivated}
\STATE{Put them into $I_F$/$I_D$ and set Dflag=True.}
\ENDIF
\ENDWHILE
\STATE{Update $t_0=t_e$.}
\ENDFOR\label{end0}
\STATE{Calculate $C_w$ from $f(t,\mathbf\Psi)$ and $w(t)$ according to \eqref{eq:TWC}. }
%\STATE{\textbf{Output:} $C_w$}
\end{algorithmic}
\end{algorithm}

\subsection{GVBS Deployment Optimization}\label{sec:GVBS}
Recall that GVBS deployment optimization is to determine the set of binary variables $\{p_{g,n}, g=1,...,G; n=1,...,N\}$, which associate $G$ GVBSs with $N$ reachable locations. In order to maximize the time-weighted coverage, the design of GVBS deployment should follow the following guidelines:
\begin{itemize}
\item{\emph{Swift deployment}: A GVBS should not be assigned to a location far-way from its original location. Specifically, we should set $p_{g,n}=0$ if the time required for the $g$th GVBS to reach the $n$th location is larger than certain threshold, denoted as $t_{th}^V$, i.e., if $t_{g,n}\geq t_{th}^V$. }
\item{\emph{Effective coverage extension}: The GVBS, once arrived, should effectively extend the coverage. In other words, the coverage area of GVBS should avoid overlapping with that of the sustained TBSs. }
\end{itemize}

 The total number of possible associations between $G$ GBVSs and $N$ locations is $\frac{N!}{(N-G)!}$.  Given the association, we can calculate the time-weighted coverage from Algorithm~\ref{alg1} (by ignoring the FBS and DBS). Hence, if $G$ and $N$ are relatively small, we can evaluate the performance of each association and return the one that gives the maximal time-weighted coverage. If  $G$ and $N$ are large, we can set a tighter time threshold, i.e., smaller $t_{th}^V$, for reducing the number of associations to be evaluated.

\subsection{FBS/DBS Deployment Optimization}
The FBS/DBS deployment is optimized with fixed TBS locations and the optimized GVBS deployment in Section~\ref{sec:GVBS}. Unlike GVBS, FBS and DBS are not constrained by the ground road condition and they can be deployed in 3D air space. Hence, their locations are continuous variables.  Furthermore,  the dispatch time of FBS and DBS must be jointly designed with the locations due to their limited endurance. Because of the sophisticated relation between the FBS/DBS deployment and the time-weighted coverage, this is still a ``black-box" optimization problem.

Among the existing techniques for handling the black-box optimization problems\footnote{Other techniques for black-box optimization include: meta learning, covariance matrix evolution, particle swarm optimization, etc\cite{TVMeta}. }, we choose GA for our problem due to its effectiveness and ability to handle large-scale problems. The dispatch time, $\{t_u^F,u=1,...,U\}$, $\{t_k^D,k=1,...,K\}$, and designated locations of FBS and DBS, $\{(x_u^F,y_u^F, u=1,...,U)\}$, $\{(x_k^F,y_k^F),k=1,...,K\}$, are jointly optimized, and hence the total number of variables is $3(U+K)$. The dispatch time of FBS and DBS are bounded within $[0,T_{th})$ and the designated locations are bounded within the disaster area, i.e., $[-R, R]$. To address the problem of different variable scales, we normalize all variables in each iteration. The GA is implemented using Matlab Genetic Algorithm toolbox.

\section{Numerical Results}
\begin{figure}[htb]
\centering
\includegraphics[scale=0.7]{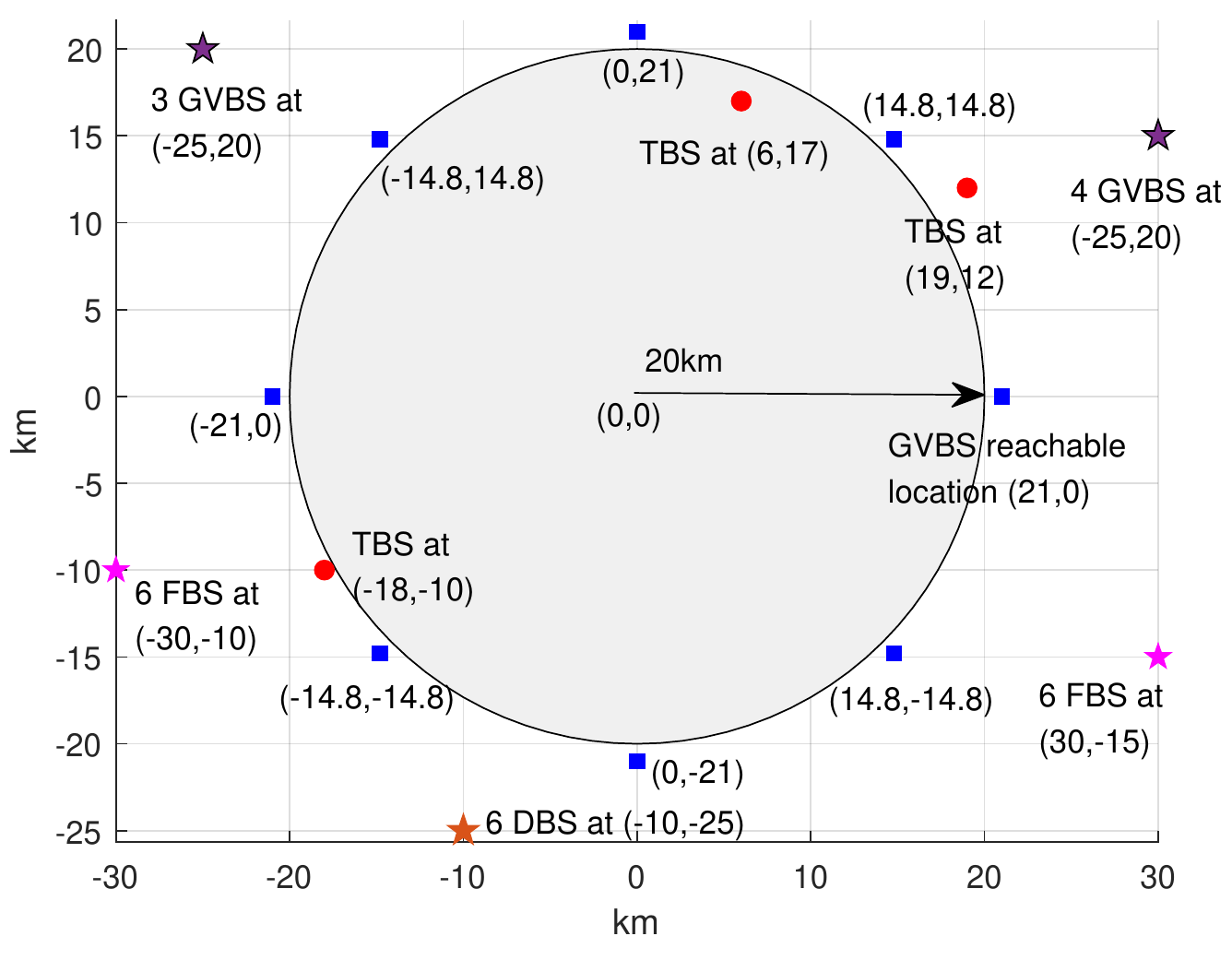}
\caption{The setup for the numerical example.}
\label{F:SimuSetup}
\vspace{-0.22in}
\end{figure}

%In this section, we evaluate the performance of the proposed integrated aerial and terrestrial communication network design using numerical examples.
We consider a disaster area of radius $20$ km. The initial locations of TBS, GVBS, FBS, DBS and the reachable locations of GVBS are shown in Fig.~\ref{F:SimuSetup}. The time required for the $g$th GVBS to reach the $n$th location is set as the required traveling time by following a straight path with an average travelling speed $30$ km/hour. Based on their different heights, we set the coverage radius of various BSs as $r_T=2$ km, $r_G=r_D=3$ km, $r_F=6$ km. The backhaul communication distance threshold between FBS and TBS, GVBS, other FBS and DBS are set as $8$ km, $8$ km, $10$ km, and $8$ km, respectively. The backhaul communication distance between DBS and TBS, GVBS, FBS and other DBS are set as $5$ km, $5$ km, $8$ km and $5$ km, respectively. The endurance of FBS and DBS are set as $2$ hours and $5$ hours, respectively. The aircrafts that carry FBS and DBS are assumed to fly at $50$ km/hour.

\begin{figure}[htb]
\centering
\includegraphics[scale=0.7]{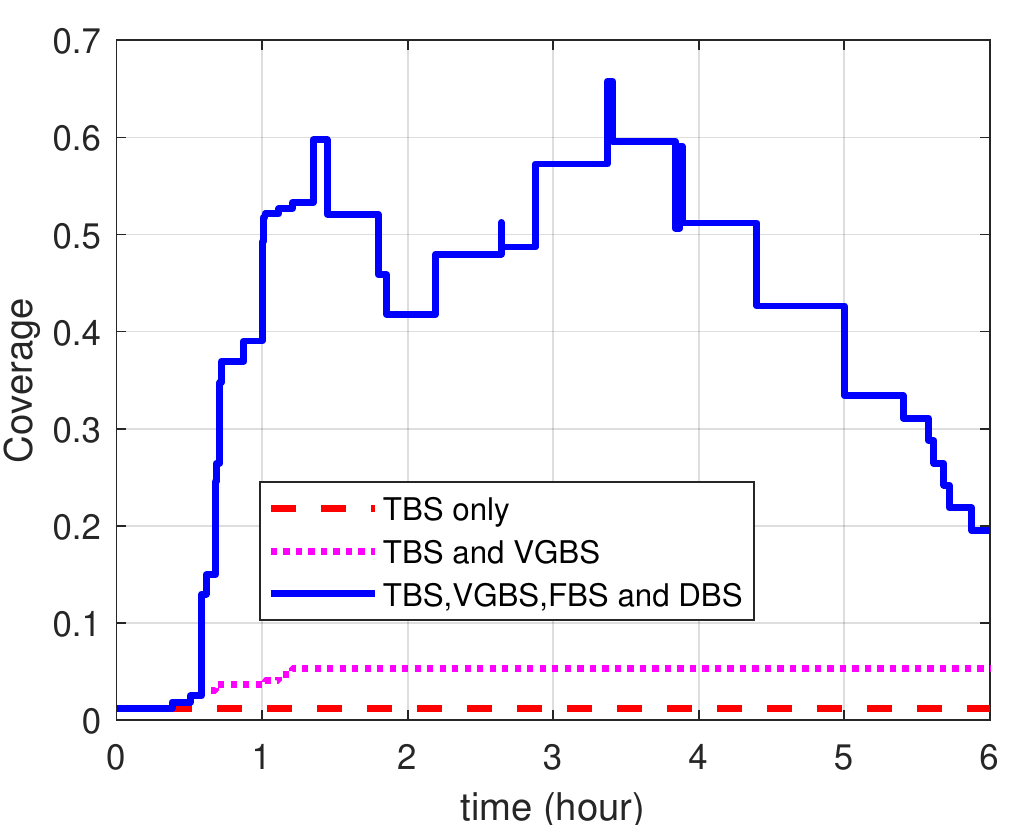}
\caption{Coverage dynamics for different network architectures.}
\label{F:compFBS}
\vspace{-0.15in}
\end{figure}

We first compare the coverage enhancement by various types of BSs. The network deployment is designed to maximize the time-weighted coverage with $w(t)=1$ and $T_{th}=5$ hours. It is observed from Fig.~\ref{F:compFBS} that the coverage enhancement by GVBS is very limited since the vehicles cannot enter the disaster center. On the other hand,  the communication coverage can be significantly enhanced by the FBS and DBS.

\begin{figure}[htb]
\centering
\includegraphics[scale=0.7]{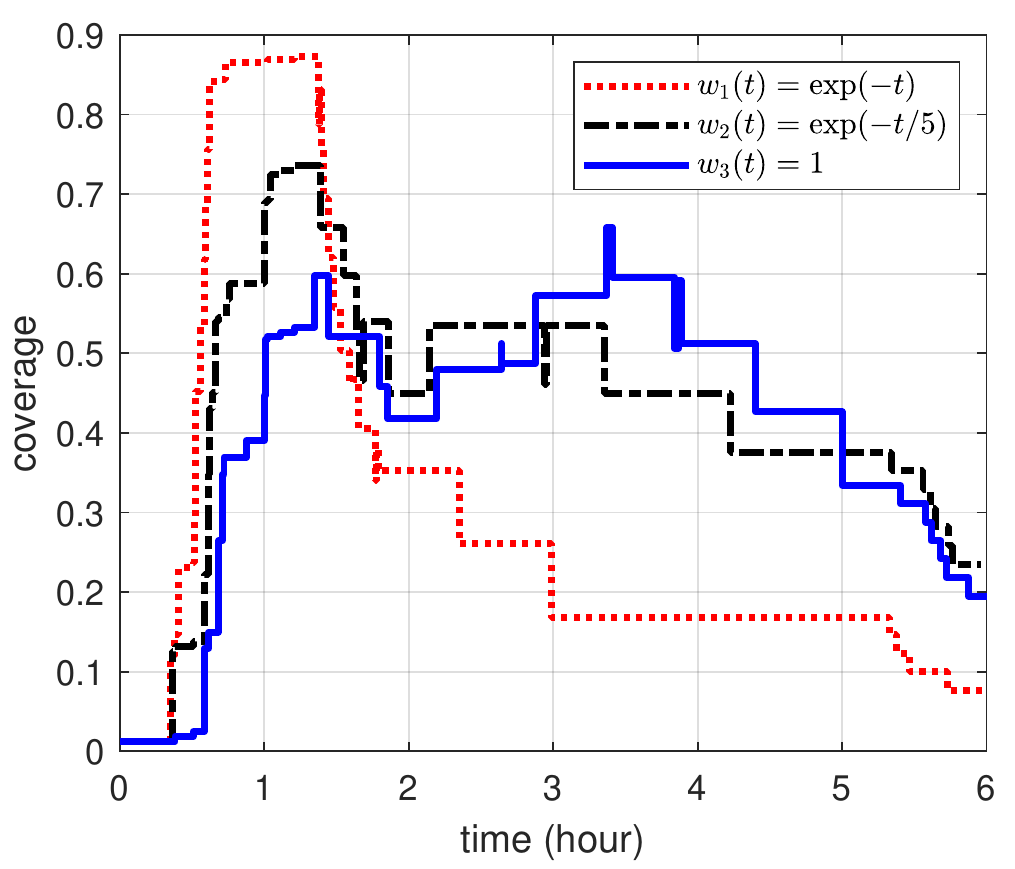}
\caption{Coverage dynamics with different weighting functions.}
\label{F:weightFunc}
\vspace{-0.15in}
\end{figure}

Next, we study the effect of weighting function on the network deployment. We use the weighting functions $w(t)=\exp\left(-\alpha t\right),\alpha\geq 0$.
%\begin{align}
%w(t)=\exp\left(-\alpha t\right), \alpha\geq 0. \label{eq:WF}
%\end{align}
When $\alpha> 0$, $w(t)$ is a decreasing function and the decreasing rate increases with $\alpha$. Fig.~\ref{F:weightFunc} compares the coverage performance for the network designs with $\alpha$ equal to $0,0.2$ and $1$, respectively. It is observed that by choosing  $\alpha=1$, the network deployment tends to achieve a wide coverage in a short time, but the coverage also shrinks quickly. On the other hand, by choosing $\alpha=0$, the network deployment tends to achieve a relatively ``stable coverage" throughout the whole considered duration. In practical scenarios, we can tune the weighting function to adjust the network design based on the requirements.

\begin{figure}[htb]
\centering
\includegraphics[scale=0.7]{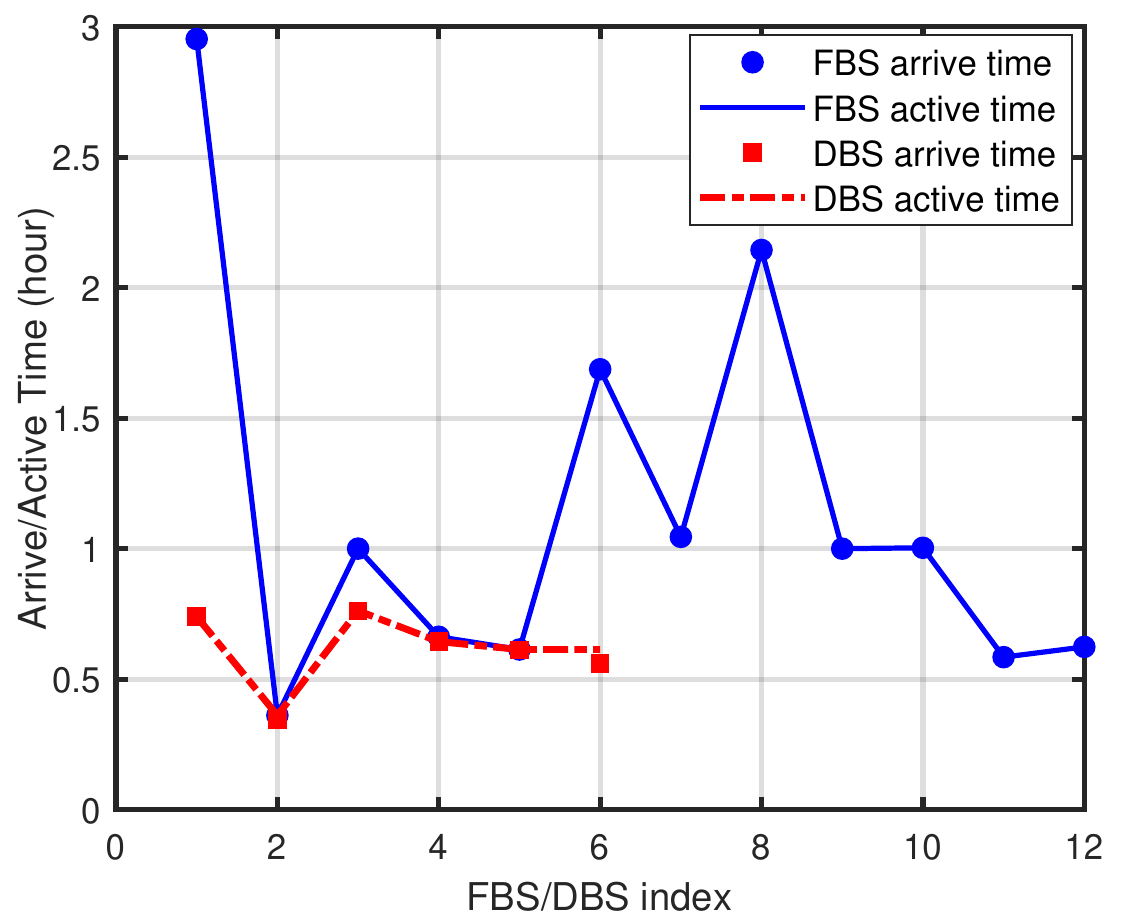}
\caption{Comparing the arriving and active time for FBS and DBS.}
\label{F:ArriveTime}
\vspace{-0.15in}
\end{figure}

Intuitively, when the endurance of the FBS/DBS is less than $T_{th}$, the optimal deployment of FBS/DBS should satisfy the following condition:
\begin{itemize}
\item{\emph{Arrive and active}: The FBS/DBS should be able to establish the backhaul connections and actively serve the ground users \emph{immediately} when they arrive at their designated locations. Otherwise, we can postpone their dispatch time to prolong their service time, and hence increase the time-weighted coverage. }
\end{itemize}
To verify the above intuition, the arrive and active time for all the FBS and DBS are compared in Fig.~\ref{F:ArriveTime} based on the obtained FBS and DBS deployment with $w(t)=\exp(-t/5)$. It is observed that most of the arrive and active times match very well, though such constraint is not imposed in GA formulation. This observation ascertains the effectiveness of the chosen optimization technique.

\begin{figure}[htb]
\centering
\includegraphics[scale=0.5]{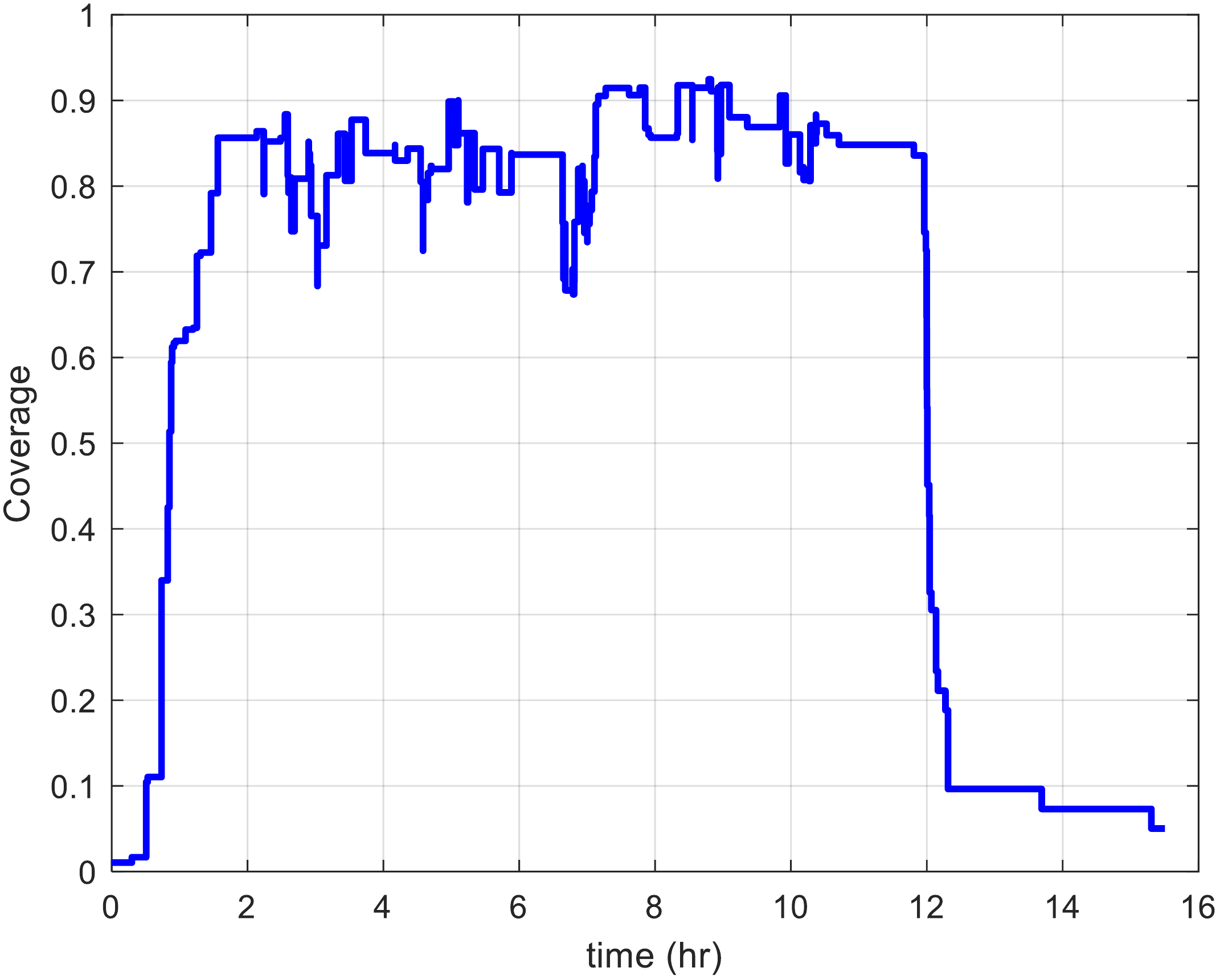}
\caption{The 12-hours coverage performance for a large network.}
\label{F:Coverage12hr}
\end{figure}

Previous numerical examples focus on the swift communication recovery for a relatively short time duration, e.g., $T_{th}=5$ hours. In practice, the FBS may continue its service after being recharged and more FBS/DBS can be delivered to the disaster area after certain time. Hence, stable communication coverage over a longer duration can be achieved. For illustration purpose, we consider the same setting as Fig.~\ref{F:SimuSetup} with the total number of FBS and DBS increased to 50 (25 at each center) and 20, respectively. The network is designed to maximize the time-weighted coverage with weighting function $w(t)=1$ and a time duration $T_{th}=12$ hours. It is observed from Fig.~\ref{F:Coverage12hr} that a stable coverage above $70\%$ can be achieved.

\section{Conclusion}
In this paper, we have proposed an integrated aerial and ground network for swift communication recovery after a catastrophic disaster. Given the locations of sustained TBS, the deployment of GVBS, FBS and DBS are jointly designed to maximize the time-weighted coverage. The results show that the FBS and DBS can significantly enhance the communication coverage since they can approach the disaster center. In the future, the dynamic network design, including UAV recharging and BS relocation process, will be considered.

\bibliographystyle{ieeetr}
\bibliography{UAV}

\end{document}